\documentclass{article} 

\usepackage{graphicx} 
\usepackage{amsmath} 
\usepackage{amssymb} 
\usepackage{setspace} 
\usepackage{hyperref} 
\usepackage{xurl} 
\hypersetup{
    colorlinks=true, 
    linkcolor=blue,  
    urlcolor=blue,   
    citecolor=blue   
}
\usepackage[a4paper,margin=1in]{geometry} 
\usepackage{microtype} 
\usepackage{titlesec} 

\usepackage[numbers, square]{natbib}  
\title{\textbf{Classical Mechanics as an Emergent Compression of Quantum Information}}
\author{
    Krzysztof Sienicki\thanks{Chair of Theoretical Physics of Naturally Intelligent Systems, Lipowa 2/Topolowa 19, 05-807 Podkowa Leśna, Poland, EU.}
}
\date{\today}

\begin{document}

\maketitle
\maketitle

\maketitle
\begin{abstract}
The correspondence principle states that classical mechanics emerges from quantum mechanics in the appropriate limits. However, beyond this heuristic rule, an information-theoretic perspective reveals that classical mechanics is a compressed, lower-information representation of quantum reality. Quantum mechanics encodes significantly more information through superposition, entanglement, and phase coherence, which are lost due to decoherence, phase averaging, and measurement, reducing the system to a classical probability distribution. This transition is quantified using Kolmogorov complexity, where classical systems require \( O(N) \) bits of information, while quantum descriptions only require $O(2^N)$, showing an exponential reduction in complexity. Further justification comes from Ehrenfest’s theorem, which ensures that quantum expectation values obey Newton’s laws, and path integral suppression, which eliminates non-classical trajectories when \( S \gg \hbar \). Thus, rather than viewing quantum mechanics as an extension of classical mechanics, we argue that classical mechanics is a lossy, computationally reduced encoding of quantum physics, emerging from a systematic loss of quantum correlations.
\end{abstract}

The relationship between classical mechanics and quantum mechanics is often understood through a heuristic rule known as the correspondence principle.\cite{bohr1920serienspektra} This rule states that classical mechanics should emerge as a special case of quantum mechanics under specific conditions. It serves as a guiding principle rather than a formal theorem, ensuring that quantum mechanics does not contradict classical physics but instead generalizes it. \cite{nielsen2013correspondence}

In essence, the correspondence principle suggests that, when quantum effects become negligible, the behavior of a system should be consistent with classical laws. This transition occurs in certain limits, such as when the action \( S \) of a system is much greater than the Planck constant \( \hbar \), leading to the suppression of quantum interference. One should notice that the standard notion of a classical limit, represented schematically by $\hbar \to 0$, provides a doubtful method for approximating a quantum system by a classical one, since Planck's constant is a fundamental constant. \cite{klein2012limit}, \cite{layton2024classical}

With increasing sizes, the quantum actions of the particles merge into average behavior, leading to the predictable motion determined by Newtonian mechanics. Another critical mechanism is decoherence, in which a quantum system in contact with the environment loses coherence through the interaction, thereby changing quantum superpositions to classical probability distributions.

The correspondence principle states that classical physics is not a separate paradigm but an oversimplified version of something that exists on a deeper level: quantum reality. Within certain bounds, there is a guarantee between classical mechanics and quantum mechanics, where the two systems predict the same outcomes. This is one of those principles that is crucial in physics and provides some form of logic or connection between the quantum actions at the microscopic level to the actions on the macroscopic scale observable at the phenomena level.

The conventional formulation of the correspondence principle considers classical mechanics to be an approximation of quantum mechanics in certain bounds. However, this view does not account for the \textit{information-theoretic aspects of this transition}. Rather than considering classical mechanics as a special case of quantum mechanics, classical mechanics can be understood as an approximation representing the lower information of quantum reality.

In quantum mechanics, a system is described by a \textit{wavefunction} \( |\psi\rangle \), which encodes \textit{amplitudes and phase information} in a high-dimensional Hilbert space. This state contains \textit{more information} than a classical description because it allows \textit{ superposition and entanglement}, which means multiple computational paths contribute to the evolution of the system.

However, in classical mechanics, a system is fully described by its \textit{position and momentum} at a given time, requiring significantly \textit{less information}. The transition from quantum to classical mechanics occurs when quantum coherence, interference, and entanglement are lost due to mechanisms such as \textit{decoherence, phase averaging, and measurement}. This effectively removes redundant quantum correlations and reduces the Kolmogorov complexity of the system description. \cite{li2008introduction}, \cite{mora2007quantum}

Mathematically, this compression can be expressed as:
\begin{equation}
    K_C(P) \ll K_Q(|\psi\rangle),
\end{equation}
where \( K_Q(|\psi\rangle) \) is the Kolmogorov complexity of the quantum state and \( K_C(P) \) is the Kolmogorov complexity of the corresponding classical probability distribution. Since the quantum description requires storing both amplitudes and phases, while the classical description only retains probabilities, classical mechanics can be seen as a lossy compression of quantum mechanics, where fine-grained quantum information is discarded.

Thus, rather than viewing classical mechanics as simply a limit of quantum mechanics, it is more accurately described as a computationally reduced encoding of quantum physics. This perspective highlights that classical mechanics is not fundamental but emerges as an effective, lower-information approximation when quantum effects become inaccessible.

Quantum mechanics encodes \textit{ much more information} than classical mechanics due to \textit{superposition (}storing multiple classical states in a single quantum state), \textit{entanglement} (reducing independent storage needs by encoding nonlocal correlations), and \textit{phase coherence} (allowing interference to process information more efficiently). When these quantum features become inaccessible, either through decoherence, phase averaging, or measurement, the resulting classical description is a coarse-grained, reduced complexity model of quantum reality. \cite{mueller2007quantum}

A fundamental way to quantify this difference is through Kolmogorov complexity, which measures the length of the shortest program capable of generating the description of a system. A classical system in phase space is specified by position \( q \) and momentum \( p \), requiring only a limited amount of information to track its trajectory. If a system has \( N \) degrees of freedom, the complexity of its classical description generally scales as  
\begin{equation}
    K_C(P) = O(N).
\end{equation}  
In contrast, a quantum system is described by a wave function \( |\psi\rangle \), which exists in a superposition of states and requires storing complex amplitudes and phase information. The complexity of a generic quantum state scales as  
\begin{equation}
    K_Q(|\psi\rangle) = O(2^N).
\end{equation}  
Since \( K_Q(|\psi\rangle) \gg K_C(P) \), a complete quantum description contains exponentially more information than its classical counterpart. The compression ratio between classical and quantum descriptions, defined as  
\begin{equation}
    R = \frac{K_C}{K_Q},
\end{equation}  
decreases exponentially with system size. In the limit \( N \to \infty \),  $\lim_{N \to \infty} R = 0$ indicating that classical descriptions provide a highly compressed and coarse-grained representation of quantum reality, omitting most quantum correlations.

This compression primarily occurs due to decoherence, which "destroys" quantum coherence by entangling a system with its environment. Mathematically, this process is described by tracing the degree of freedom of the environment, transforming a quantum state from a pure state to a classical probability distribution. A fully quantum system is represented by its \textit{density matrix}:
\begin{equation}
    \rho_{\text{quantum}} = \sum_{i,j} c_i c_j^* |x_i\rangle \langle x_j|.
\end{equation}
The off-diagonal terms \( c_i c_j^* \) encode quantum interference and coherence. When the system interacts with an environment, decoherence removes these terms, leaving only a diagonal classical probability distribution:
\begin{equation}
    \rho_{\text{classical}} = \sum_i |c_i|^2 |x_i\rangle \langle x_i|.
\end{equation}
At this point, the system can be fully described by classical probabilities \( P(x_i) = |c_i|^2 \), and the quantum coherence is lost. This transformation confirms that the classical world emerges when quantum correlations are irreversibly discarded, reducing Kolmogorov complexity from \( O(N) \) to \(O(2^N) \).

Another mechanism driving this compression is phase averaging in the Feynman-path integral formalism. \cite{feynman1965path} In quantum mechanics, the \textit{propagator} is given by the sum over all possible paths, weighted by \( e^{iS/\hbar} \), where \( S \) is the classical action:
\begin{equation}
    K(x,t) = \int \mathcal{D}[q(t)] e^{i S[q]/\hbar}.
\end{equation}
However, when \( S \gg \hbar \), rapid oscillations in \( e^{iS/\hbar} \) cause destructive interference for all non-classical paths, ensuring that only the \textit{classical trajectories remain dominant}. Mathematically, this is expressed as:
\begin{equation}
    \lim_{S \gg \hbar} K(x,t) \approx e^{i S_{\text{cl}}/\hbar}.
\end{equation}
This confirms that classical mechanics emerges as a phase-averaged approximation of quantum evolution, where only stationary-phase paths contribute.

Even within quantum mechanics, the expectation values of the observables obey the classical equations of motion. Ehrenfest’s theorem \cite{plastino1993tsallis} states that for any observable \( \hat{A} \), its expectation value evolves according to:
\begin{equation}
    \frac{d}{dt} \langle \hat{A} \rangle = \frac{i}{\hbar} \langle [H, \hat{A}] \rangle.
\end{equation}
For position and momentum, this leads to:
\begin{equation}
    \frac{d}{dt} \langle \hat{x} \rangle = \frac{\langle \hat{p} \rangle}{m}, \quad \frac{d}{dt} \langle \hat{p} \rangle = -\left\langle \frac{dV}{dx} \right\rangle.
\end{equation}
These are precisely Hamilton's classical equations, which means that classical mechanics is a \textit{ mean-field approximation of quantum dynamics} in the limit where quantum fluctuations become negligible.

The shift from quantum mechanics to classical mechanics can be thought of as a form of lossy data compression from the standpoint of information theory. A quantum state retains its information in the form of \textit{entanglement}, which provides a system with easy global encoding. On the other hand, classical mechanics does not have entanglement, and hence it has to store independent state variables for every degree of freedom, which is computationally wasteful. The classical description is more elaborate than its quantum counterpart because it employs more classical states than the nonlocal correlations in quantum mechanics. So the transition from the quantum world to the classical world can be viewed as an optimization process for storage, where the requirement of tracking full quantum correlations comes with a computational price for the system: destruction of phase coherence.   

Yet, examining the quantum-to-classical transition from the perspective of Kolmogorov complexity, decoherence and phase averaging reveals an information-theoretic facet of classical emergence. Different yet complementary to Kolmogorov complexity is Bell’s theorem regarding the existence of nonlocal features without a classical equivalent, and discussing the quantum realm.

Bell’s theorem states that any local hidden variable theory cannot reproduce the statistical predictions of quantum mechanics. \cite{zych2019bell} More formally, Bell’s inequalities impose a constraint on classical correlations that is violated by quantum entanglement, demonstrating that quantum mechanics cannot be simulated by any classical theory based on local realism.  

In the context of the correspondence principle (see above) and information compression, this raises an important question: If classical mechanics emerges from quantum mechanics as a compressed representation, why does classical physics lack non-local correlations?  

The answer lies in decoherence and environment-induced information loss. Bell’s theorem does not state that non-locality disappears in macroscopic systems, it simply shows that classical theories cannot account for it. However, when a quantum system undergoes decoherence, entangled states lose coherence and become effectively separable, transforming quantum correlations into classical probability distributions (see also eqs. (5) and (6) above) :
\begin{equation}
    \rho_{\text{entangled}} = \sum_{i,j} c_i c_j^* |x_i\rangle \langle x_j| \quad \longrightarrow \quad \rho_{\text{classical}} = \sum_i |c_i|^2 |x_i\rangle \langle x_i|.
\end{equation}
This transformation removes violations of Bell’s inequality at macroscopic scales. Thus, Bell’s theorem highlights a key feature of quantum mechanics: entanglement cannot be described classically, while the correspondence principle explains how classical mechanics emerges when this entanglement is lost.  
 
Bell’s theorem and the correspondence principle describe two complementary aspects of quantum-classical emergence. Bell’s theorem shows that quantum mechanics cannot be simulated by local hidden variable theories, highlighting the uniquely nonlocal nature of quantum correlations. The correspondence principle and information compression explain how the classical world emerges from quantum reality by discarding nonlocal quantum correlations through decoherence and complexity reduction. Thus, the classical world does not contradict Bell's theorem, but emerges as an effective computationally reduced approximation where quantum nonlocality is no longer accessible. This perspective reinforces the idea that classical mechanics is a lossy encoding of quantum information, with decoherence acting as the mechanism that suppresses violations of Bell’s inequality at macroscopic scales.

The key conclusion is that classical mechanics is not a fundamental theory, but a computationally reduced encoding of quantum mechanics. This transition occurs when entanglement and phase information are removed, either due to environmental interactions (decoherence) or due to coarse-graining in high-action regimes. The loss of quantum coherence leads to a significant reduction in Kolmogorov complexity, making classical mechanics a low-information approximation of the full quantum description.

Rather than viewing quantum mechanics as an extension of classical mechanics, we should view classical mechanics as an information-theoretically degraded version of quantum mechanics. Classical physics is what remains when quantum information is compressed, phase-averaged, and decohered away. The classical world is what remains when quantum information is compressed, decohered, and phase averaged away.

\textbf{Declaration}: the author declares no conflicts of interest. The author used ChatGPT, Gemini, Copilot and/or Grammarly to translate the article from Polish and/or to refine the English language, if at all. However, the author assumes full responsibility for all errors and shortcomings.  

\bibliography{ReferencesQMCM} 
\end{document}